\documentclass[a4paper,UKenglish,cleveref, autoref, thm-restate]{lipics-v2021}
\usepackage[T1]{fontenc}
\usepackage{graphicx}
\usepackage{hyperref}
\usepackage{amsmath}
\usepackage[noabbrev,capitalise]{cleveref}
\usepackage{url}
\usepackage{svg}
\usepackage{wrapfig}

\usepackage{tikz,pgfplots}

\definecolor{my-dark-red}{RGB}{183, 28, 28}
\definecolor{my-red}{RGB}{244,67,54}
\definecolor{my-pink}{RGB}{233,30,99}
\definecolor{my-purple}{RGB}{156,39,176}
\definecolor{my-deep-puple}{RGB}{103,58,183}
\definecolor{my-indigo}{RGB}{63,81,181}
\definecolor{my-blue}{RGB}{33,150,243}
\definecolor{my-light-blue}{RGB}{3,169,244}
\definecolor{my-cyan}{RGB}{0,188,212}
\definecolor{my-teal}{RGB}{0,150,136}
\definecolor{my-green}{RGB}{76,175,80}
\definecolor{my-light-green}{RGB}{139,195,74}
\definecolor{my-lime}{RGB}{205,220,57}
\definecolor{my-yellow}{RGB}{255,235,59}
\definecolor{my-amber}{RGB}{255,193,7}
\definecolor{my-orange}{RGB}{255,152,0}
\definecolor{my-deep-orange}{RGB}{255,87,34}
\definecolor{my-brown}{RGB}{121,85,72}
\definecolor{my-grey}{RGB}{158,158,158}
\definecolor{my-dark-grey}{RGB}{111,111,111}
\definecolor{my-blue-grey}{RGB}{96,125,139}

\pgfplotscreateplotcyclelist{mycolor}{%
  my-red, every mark/.append style={solid,scale=0.8}, mark=x \\%
  my-blue, every mark/.append style={solid,scale=0.6}, mark=o \\%
  my-light-green, every mark/.append style={solid,scale=0.8}, mark=diamond \\%
  my-purple, every mark/.append style={solid,scale=0.8,rotate=180}, mark=triangle \\%
  my-teal, every mark/.append style={solid,scale=0.65}, mark=square \\%
  my-amber, every mark/.append style={solid,scale=0.8}, mark=star \\%
  my-deep-orange, every mark/.append style={solid,scale=0.8}, mark=triangle \\%
  my-indigo, every mark/.append style={solid,scale=0.8}, mark=+ \\%
  my-orange, every mark/.append style={solid,scale=0.8}, mark=pentagon \\%
  my-brown, every mark/.append style={solid,scale=0.9}, mark=Mercedes star flipped \\%
  my-red, every mark/.append style={solid,scale=0.8}, mark=x, dashed \\%
  my-blue, every mark/.append style={solid,scale=0.6}, mark=o, dashed \\%
  my-light-green, every mark/.append style={solid,scale=0.8}, mark=diamond, dashed \\%
  my-purple, every mark/.append style={solid,scale=0.8,rotate=180}, mark=triangle, dashed \\%
  my-teal, every mark/.append style={solid,scale=0.65}, mark=square, dashed \\%
  my-amber, every mark/.append style={solid,scale=0.8}, mark=star, dashed \\%
  my-deep-orange, every mark/.append style={solid,scale=0.8}, mark=triangle, dashed \\%
  my-indigo, every mark/.append style={solid,scale=0.8}, mark=+, dashed \\%
  my-organge, every mark/.append style={solid,scale=0.8}, mark=pentagon, dashed \\%
  my-brown, every mark/.append style={solid,scale=0.9}, mark=Mercedes star flipped, dashed \\%
  my-red, every mark/.append style={solid,scale=0.8}, mark=x, densely dotted \\%
  my-blue, every mark/.append style={solid,scale=0.6}, mark=o, densely dotted \\%
  my-light-green, every mark/.append style={solid,scale=0.8}, mark=diamond, densely dotted \\%
  my-purple, every mark/.append style={solid,scale=0.8,rotate=180}, mark=triangle, densely dotted \\%
  my-teal, every mark/.append style={solid,scale=0.65}, mark=square, densely dotted \\%
  my-amber, every mark/.append style={solid,scale=0.8}, mark=star, densely dotted \\%
  my-deep-orange, every mark/.append style={solid,scale=0.8}, mark=triangle, densely dotted \\%
  my-indigo, every mark/.append style={solid,scale=0.8}, mark=+, densely dotted \\%
  my-orange, every mark/.append style={solid,scale=0.8}, mark=pentagon, densely dotted \\%
  my-brown, every mark/.append style={solid,scale=0.9}, mark=Mercedes star flipped, densely dotted \\%
}

\pgfplotsset{
  major grid style={thin,dotted,color=my-blue-grey!80!black},
  minor grid style={thin,dotted,color=my-grey},
  ymajorgrids,
  yminorgrids,
  cycle list name={mycolor},
  every axis/.append style={
    line width=0.5pt,
    tick style={
      line cap=round,
      thin,
      major tick length=4pt,
      minor tick length=2pt,
    },
  },
  legend cell align=left,
  legend style={
    /tikz/every even column/.append style={column sep=3mm,black},
    /tikz/every odd column/.append style={black},
  },
  legend style={font=\small},
  title style={yshift=-2pt},
  enlarge x limits=0.04,
  every tick label/.append style={font=\footnotesize},
  every axis label/.append style={font=\small},
  every axis y label/.append style={yshift=-1ex},
  plotSelectQueryTimeAll/.style={
    axis lines=left,
    x axis line style={-},
    y axis line style={-},
    axis line shift=10pt,
    ymajorgrids=true,
    grid=both,
    width=44.0mm,
    height=48.5mm,
    ymin=50,
    ymax=375,
  },
  plotRankQueryTimeAll/.style={
    axis lines=left,
    x axis line style={-},
    y axis line style={-},
    axis line shift=10pt, 
    ymajorgrids=true,
    grid=both,
    width=44.0mm,
    height=48.5mm,
    ymin=30,
    ymax=70,
  },
  plotConstructionTimeAll/.style={
    axis lines=left,
    x axis line style={-},
    y axis line style={-},
    axis line shift=10pt,
    ymajorgrids=true,
    xmajorgrids=true,
    grid=both,
    width=44.0mm,
    height=50.0mm,
    ymin=0,
    ymax=220,
  },
  plotMemoryRequirementsAll/.style={
    axis lines=left,
    x axis line style={-},
    y axis line style={-},
    axis line shift=10pt,
    ymajorgrids=true,
    minor x tick num=1,
    grid=both,
    xlabel near ticks,
    ylabel near ticks,
    tick align=outside,
    width=50.0mm,
    height=50.0mm,
    ymin=100,
    ymax=450,
  }
}
\usepackage{booktabs}
\usepackage{listings}

\hypersetup{
  colorlinks=true,
  pdftitle={Engineering Compact Data Structures for Rank and Select Queries on Bit Vectors},
  pdfauthor={Florian Kurpicz},
  pdfsubject={}
}

\title{Engineering Compact Data Structures for Rank and Select Queries on Bit Vectors} 

\author{Florian Kurpicz}{Karlsruhe Institute of Technology, Germany \and \url{https://kurpicz.org}}{kurpicz@kit.edu}{https://orcid.org/0000-0002-2379-9455}{}

\authorrunning{F. Kurpicz} 

\Copyright{Florian Kurpicz} 



\category{} 

\relatedversion{} 

\supplement{All implementations presented in this paper and scripts to reproduce our experimental evaluation are available under the GPLv3 license.}
\supplementdetails[]{Rank and select data structure implementations}{https://github.com/pasta-toolbox/bit_vector}
\supplementdetails[]{Scripts for reproduction of results}{https://github.com/pasta-toolbox/bit_vector_experiments}

\nolinenumbers 

\ArticleNo{1}

\begin{document}

\maketitle              

\begin{abstract}
  Bit vectors are fundamental building blocks of succinct data structures used in compressed text indices, e.g., in the form of the wavelet trees.
  Here, two types of queries are of interest: rank and select queries.
  In practice, the smallest (uncompressed) rank and select data structure cs-poppy has a space overhead of \(\approx\) 3.51\,\% [Zhou et al., SEA~‘13].
  Using the same overhead, we present a data structure that can answer queries up to 8\,\% (rank) and 16.5\,\% (select) faster compared with cs-poppy.
\end{abstract}

\section{Introduction and Related Work}
Given a bit vector \textsf{B} of length \(n\) and \(\alpha\in\{0,1\}\), \emph{rank} and \emph{select} are defined as:
\begin{description}
\item[rank:] given \(i\in[0,n)\), rank returns the number of ones (or zeros) in \(\textsf{B}[0,i]\), i.e., \[\textsf{B}.rank_\alpha(i)=|\{j\in[0,i]\colon \textsf{B}[j]=\alpha\}|\]
\item[select:] given a rank \(i\), select returns the leftmost position where the bit vector contains a one (or zero) with rank \(i\), i.e.,  \[\textsf{B}.select_\alpha(i)=\min\{j\in[0,n)\colon \textsf{B}.rank_\alpha(j)=i\}\]
\end{description}

Bit vectors are building blocks of many important compact and succinct data structures like wavelet trees \cite{FerraginaM2000FMIndex} that have applications in many compressed full-text indices (e.g., the FM-index~\cite{FerraginaM2000FMIndex} and \(r\)-index~\cite{GagieNP2020FullyFunctionalRIndex}; we point to the following surveys \cite{DinklageEFKL2021PracticalWaveletTrees,FerraginaGM2009MyriadWT,Makris2012WaveletSurvey,Navarro2014WaveletForAll} for more information on wavelet trees), succinct graph representations (e.g., LOUDS \cite{Jacobson1989LOUDS}), and can also be used as a representation of monotonic sequences of integers (e.g., Elias-Fano coding \cite{Elias1974EliasFano,Fano1971EliasFano}) that supports predecessor queries.
It should be noted that all of the applications mentioned above require rank and/or select queries on bit vectors.

Given a length-\(n\) bit vector, it is known how to solve rank and select queries in constant time using only \(n+o(n)\) bits of space~\cite{ClarkM1996Select,Jacobson1989LOUDS}.
Here, the bit vector occupies \(n\) bits.
The rank and select data structures require only \(o(n)\) additional bits.
There also exist more precise results, when considering a length-\(n\) bit vector containing \(k\) ones, focusing on applications where the number of ones is small.
Currently, the best known result requires \(\lg\binom{n}{k}+\frac{n}{(\lg n/t)^t}+\tilde O(n^{3/4})\) bits of space and can answer rank and select queries in \(O(t)\) time~\cite{Patrascu2008Succincter} (by no explicitly storing the bit vector).

\paragraph*{Related Work.}
In this paper, we focus on practical space-efficient \emph{uncompressed} rank and select data structures that can handle bit vectors of arbitrary size.
Prominent implementations can be found in the popular succinct data structure library (SDSL)~\cite{GogBMP2014SDSL}.
Furthermore, there exist highly tuned select-only data structures by Vigna~\cite{Vigna2008BroadwordRankSelect}, which currently can answer select queries the fastest while being reasonably space-efficient.
The currently most space-efficient rank and select data structure by Zhou~et~al.~\cite{ZhouAK2013PopcountRankSelect} requires only 3.51\,\% additional space.
There exists more work on practical space-efficient rank and select data structures for bit vectors that require more space, answer queries slower, and/or can handle only bit vectors up to size \(2^{32}\), e.g., \cite{GonzalezGMN2005PracticalRankSelect,KimNKP2005RankAndSelect,NavarroP2012CombinedSampling}, which we, therefore, do not consider in this paper.
There also exists work on \emph{compressed} rank and select data structures, e.g., \cite{ArroyueloW2020CompressedRankSelect,BeskersF2014CompressedRankSelect,BoffaFV2022LearnedRankSelect,RamanRS2007RRR}
and on rank and select data structures for \emph{mutable} bit vectors, e.g., \cite{PibiriK2021MutableRankSelect,Prezza2017DYNAMIC}.

\section{Preliminaries}
Due to the simplicity of the problem, the notations in this paper are rather simple.
We have a \emph{bit vector} of length \(n\), where we can access each bit in constant time.
In the following descriptions, we make use of the notion of \emph{blocks}.
Here, a block is an interval within the bit vector that \emph{covers} the bits within the interval.
Given (a part of) a bit vector, the \emph{population count} or \emph{popcount} is the number of ones within (the part of) the bit vector.
Popcount instructions are supported by most modern CPUs for up to 64 bit words.
\label{sec:different_rank_and_select_structures}
Since their first introduction, rank and select data structures that require sub-linear additional space and can answer rank and/or select queries in constant time have similar structures.
In this section, we briefly describe the commonly used designs for rank and select data structures.

Almost all practical rank data structures follow the same layout.
First, the bit vector is partitioned into consecutive \emph{basic blocks}, which are the smallest unit for which any information is collected in an index.
The rank of bits within a basic block is determined directly on the bit vector.
Then, there exists a hierarchy of different (overlaying) blocks of different sizes.
For each type of block, there exists an index that stores information about the number of ones.
The scope in which the number of ones is considered can differ, e.g., the number of ones in the block or the number of ones up to the beginning of the block from the beginning of either the bit vector or another overlaying block.
See \cref{fig:example_cs_poppy} for an example.
Queries are then answered using the information provided by the blocks and the popcount of the basic block up to the position.
Depending on the sizes of the blocks, the pertinent information in the indices can be accessed very efficiently.
It should be noted, that it suffices to store information about the number of ones in each block, as \(rank_0(i)= i-rank_1(i)\).

Unlike rank data structures, select data structures come in two flavors: \emph{rank}-based and \emph{position}-based.
Rank-based select data structures utilize a rank data structure that is enhanced by a small index containing sample positions for every \(k\)-th one (or zero).
To answer a query, we first determine the closest block using the sampled positions.
Then, we have to look at blocks until we have found the basic block that contains the correct bit with the requested rank.
The position in the basic block can then easily be computed.
While fast in practice (see \cref{sec:experimental_evaluation}), this type of select data structure usually cannot guarantee a constant query time.

Position-based select data structures on the other hand only store sample positions.
Usually, they differentiate between different block sizes and densities, e.g., for very sparse blocks, the answer of every select query can be stored directly.
One advantage of position-based select data structures is a constant worst-case query time can be achieved, e.g.,~\cite{Vigna2008BroadwordRankSelect}.
However, there is also a disadvantage of position-based select data structures.
Unlike when answering rank queries, we cannot use a \(select_1\) query to answer a \(select_0\) query.
The significant difference between rank- and position-based select data structures is that we can easily use the information in a rank-based data structure to answer both \(select_0\) and \(select_1\) queries.
We can simply transform the number of ones (or zeros) up to a position to the number of zeros (or ones) up to that position.\footnote{The sampled positions described above have to be stored for ones and zeros.}
The same is not possible with position-based select data structures.

\section{Space Efficient Rank and Select Data Structures}
\label{sec:space_efficient_rank_and_select}
First, in \cref{sec:cs-poppy}, we describe the design of a space efficient rank and select data structure by Zhou et al.~\cite{ZhouAK2013PopcountRankSelect} named \emph{cs-poppy} that makes use of fast popcount instructions.
In \cref{sec:flat_popcount}, we present our main result---a significantly faster rank and select data structure---that requires the same space as cs-poppy and makes use of SIMD.
Finally, in \cref{sec:wide_popcount}, we present a new and relatively simple rank data structure that provides better rank query times (in practice) by combining some techniques from our main result with a slightly higher space usage.

\subsection{CS-Poppy: Rank-Based Rank and Select Data Structure}
\label{sec:cs-poppy}
\begin{figure}[t]
  \centering
  \includesvg[scale=.75]{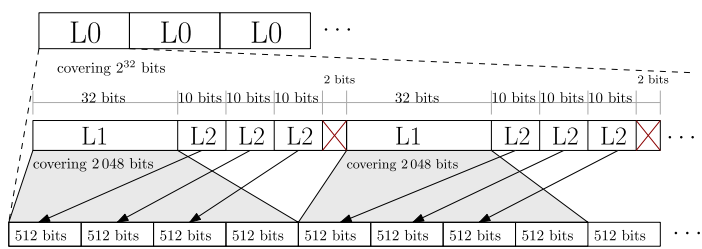}
  \caption{L0-index and interleaved L1- and L2-index of cs-poppy. Arrows indicate that the popcount of the basic block is stored.
  Wasted bits are marked by red crosses.}
  \label{fig:example_cs_poppy}
\end{figure}

We describe the rank and select data structure \emph{cs-poppy} top-down, starting with the largest blocks.
For an overview, see \cref{fig:example_cs_poppy}.
First, the bit vector is split up into consecutive L0-blocks of size \(2^{32}\) bits.
For each L0-block we store the number of ones occurring in the bit vector before the L0-block in an L0-index.
To accommodate bit vectors of arbitrary size, each entry in the L0-index requires 64 bits.
Note that the indices for the different block types are just plain arrays where the entry of the \(i\)-th block is stored at the \(i\)-th array entry.
Given a position in the bit vector, we can identify all blocks that cover the position by dividing it by the block size (in bits).
Next, each L0-block is split up into consecutive L1-blocks of size 2048 bits.
This time, we are interested in the number of ones occurring in the L0-block before the L1-block.
For each L1-block, we store this information in the L1-index.
Since the number of ones in an L0-block can be at most \(2^{32}\), each entry in the L1-index requires only 32 bits.
Finally, each L1-block is split up into four L2-blocks of size 512, i.e., the basic blocks of cs-poppy.
We store the number of ones in the L2-blocks for the first three L2-blocks in each L1-block in the L2-index.
The number of ones in each L1-block's last L2-block is not stored, as it does not provide any information that cannot be computed using the L1-index.
(We can look at the number of one-bits occurring before the next L1-block and subtract the number of one-bits seen before the fourth L2-block.)
A L2-index entry has to encode a number in \([0,512]\) and thus requires 10 bits.

One important technique used by the authors of cs-poppy is the \emph{interleaved L1- and L2-index}.
Here, for each L1-block and the corresponding L2-blocks, a 64-bit word is used to store the entry of the L1- and L2-index.
Since the L1-index contains 32-bit words and the L2-index contains 10-bit entries (of which three are pertinent to the L1-block), everything fits into 64 bits.
While this approach wastes two bits for each L2-block (0.09\,\% additional space), it reduces the number of cache misses, as the required part of the L2-index should be loaded whenever the L1-index is accessed.
Zhou et al.~\cite{ZhouAK2013PopcountRankSelect} introduced more practical improvements that can speed up answering rank and select queries using cs-poppy.
We refer to their paper for a detailed description.

\paragraph*{Answering Rank Queries.}
Now, we want to answer a rank query for position \(i\).
To this end, we first have to identify the L0- and L1-block the position is covered by.
We obtain both blocks by dividing \(i\) with the bit size of an L0- and L1-block respectively.
The corresponding entries in the L0- and L1-index contain the number of bits occurring from the beginning of the bit vector to the beginning of the L0-block and from the beginning of the L0-block to the beginning of the L1-block.
This is the first part of the result.
Next, we have to determine the number of ones within the L1-block up to the position.
To this end, we scan the entries of the L2-index pertinent to the L1-block until we have reached the L2-block that covers position \(i\).
We add entries of the L2-index we have scanned to the result.
Finally, we have to determine the number of ones in the final L2-block directly on the bit vector.
This can be done using fast popcount instructions.
Overall, rank queries can be answered in constant time using cs-poppy.

\paragraph*{Answering Select Queries.}
To answer a select query for a rank \(i\), we first identify the L0-block where the first position with rank \(i\) can occur.
This can be done by scanning the L0-index until we see an entry greater or equal to \(i\).
The position has to be in the L0-block belonging to the previous entry in the L0-index.
Now, we have to scan the L1-index, until we have identified the L1-block that contains the searched position.
To speed up select queries, the position of every \(8\,192\)-th one is sampled.
We can use these samples to skip some parts of the L1-index.
When we have identified the correct L1-block, we continue scanning the L2-index until we find the L2-block that contains the position we are looking for.
During each step, we subtract the number of ones in the L0-, L1-, and L2-entries from the rank \(i\), because otherwise, we could not identify the block containing the result in the L1- and L2-index, resp.
Finally, we have to identify the position within the L2-block and return it.
The search in the L2-block has been further optimized by using SIMD by Pandey~et~al.~\cite{PandeyBJ2017FastSelect}.
Note that this query algorithm requires linear time in the worst-case, but is fast in practice, see \cref{sec:experimental_evaluation}.

\subsection{Flat-Popcount: Storing More Information Wasting No Bits}
\label{sec:flat_popcount}
Now, we present the main result of this paper, a rank and select data structure that requires the same space as cs-poppy but is faster in practice, see \cref{sec:experimental_evaluation}.
We call this data structure \emph{flat-popcount}. 
To achieve this, we have to store additional information about blocks without using any additional space.
Here, we make use of the two bits that cs-poppy wastes for every L1-block.

\begin{figure}[t]
  \centering
  \includesvg[scale=.75]{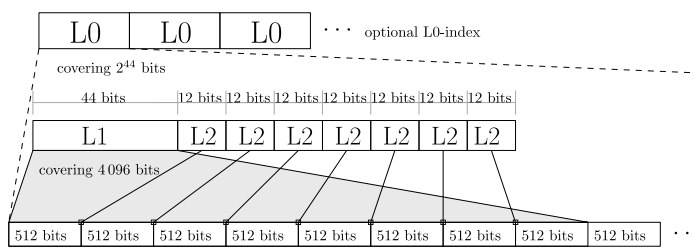}
  \caption{L0-index and interleaved L1- and L2-index of flat-popcount. Boxes indicate that the number of ones in the L1-block up to this point is stored in the L2-index.}
  \label{fig:example_flat_popcount}
\end{figure}

As mentioned before, we want to store additional information.
To be more precise, we store the number of ones that occur before each L2-block within each L1-block---similar to the L0- and L1-index---see \cref{fig:example_flat_popcount} for an example.
If we consider cs-poppy, we have to store numbers in \([0,1536]\), which require 11 bits each.
Thus, there is not enough space to do so using only 32 bits.
Our solution to this problem is to double the size of an L1-block.
We still have L2-blocks that cover consecutive 512 bits of the bit vector.
Therefore, we now have to consider eight L2-blocks within one L1-block, i.e., each L1-block covers 4\,096 bits (compared with 2\,048 bits in cs-poppy).

On the other hand, we now have 128 bits to store any information regarding the L1- and eight L2-blocks, i.e., two times 64 bits used in cs-poppy.
As before, we do not have to store any information regarding the last L2-block in the L1-block, as we can compute this information using the information stored for the next L1-block.
We can therefore store the number of ones up to each L2-block within the L1-block using 12 bits each, as the number of previously set bits is in \([0,3\,584]\).
This requires \(7\cdot 12=84\) bits, leaving us with 44 bits for the entry in the L1-index.
Similar to cs-poppy, we interleave the L1- and L2-index within one 128-bit word, which is supported by almost all modern CPUs.
Not only can we now store more information for each L2-block, we can also increase the size of the L0-blocks to \(2^{44}\) bits.
This allows us to make the L0-index optional, as it would be required only for bit vectors of size greater than \(2^{44}\) bits, which is significantly larger than the \(2^{32}\) bits that cs-poppy supports without L0-index and would require 2\,TiB space for the bit vector alone.

The space requirements of both cs-poppy and flat-popcount are nearly the same.
However, there is one big advantage of our approach: in flat-popcount, we have random access to the L2-index.
When using cs-poppy, we have to scan the entries in the L2-index, as they are delta encoded.
Using flat-popcount, we store the number of ones occurring to the left of the L2-block within the L1-block \emph{directly}, as for the L1- and L0-index.
This allows us to answer both rank queries slightly and select queries significantly more efficiently (in practice).
We now take a look at the changes in the query algorithms.

\subsubsection{Answering Rank Queries with Flat-Popcount.}
Answering rank queries for a position \(i\) in this data structure is similar to answering rank queries using cs-poppy, which we describe in \cref{sec:cs-poppy}.
At least for identifying the entries in the L0- and L1-index.
Here, nothing has changed, as we only improved on the L2-index.
As mentioned earlier, the main advantage of our new design is that we do not have to scan all L2-entries to compute the number of one-bits up the final L2-block.
Thus, we can now determine the number of bits occurring in the L1-block before the L2-block that contains the position \(i\) accessing only one entry of the L2-index.
This entry can be computed the same way we compute the entries of the L0- and L1-index.
In our experiments (see \cref{sec:experimental_evaluation}), we can see that scanning up to three L2-entries and adding up their values contributes significantly to the running time of a query.

\subsubsection{Answering Select Queries with Flat-Popcount.}
Answering a select query for a rank \(i\) is also very similar to cs-poppy.
We still sample the position of every \(8\,192\)-th one in the bit vector and use the samples to speed up identifying the L1-block that contains the first position that has rank \(i\).
Thus, the first part of the query algorithm that changes is the identification of the L2-block, where the first position with rank \(i\) occurs.
Due to us storing the number of ones occurring (within the L1-block) before each L2-block, we have a monotonic increasing sequence of integers, allowing us to search for the L2-block more efficiently.

\paragraph*{Finding the Correct L2-Entry using Linear or Binary Search.}
When we use a linear search, we scan the L2-index until we have found an entry that contains the rank that we are looking for.
Therefore, answering queries is not much different than before, when using cs-poppy.
However, since we do store the number of ones before the L2-block, we save some additions (of L2-entries) when answering queries, which results in a small improvement.
Since the number of ones in the L1-block occurring before each L2-block is a monotonic increasing sequence, we can also use a binary search on the entries of the L2-index.
As we store seven entries in the L2-index for each L1-block, we can use a \emph{uniform} binary search \cite[p.~414f]{Knuth1998TAOCPIII}, which always requires three iterations to identify the correct block.
Most importantly, we can pre-compute the whole decision tree and reuse entries of the L2-index.
Both of these approaches are rather simple and do not make use of the fact that all entries of the L2-block that we are interested in are contained in a single 128-bit word.
Modern CPUs are able to compare multiple values contained in such a word at the same time.
However, there are some limitations, which we describe in the following section.

\paragraph*{Finding the Correct L2-Entry using SIMD.}
In addition to these more obvious approaches described above, we can also search for the L2-block using the \emph{streaming SIMD extension} (SSE), which allows us to divide a 128-bit computer word into consecutive blocks and apply operations on each of the blocks at the same time.
One limitation of these instructions is that all blocks must have the same size and that the sizes of the block must be either 8, 16, 32, or 64 bits.
Our main goal is to use the compare operation \texttt{\_mm\_cmpgt\_epi16} to compare all seven entries of the L2-index that are covered by the L1-block at the same time.
Unfortunately, we cannot simply use the compare operation, because we have to store the L2-entries using only 12 bits and there is no compare operation working on 12-bit blocks.

\begin{figure}[t]
  \centering
  \includesvg[]{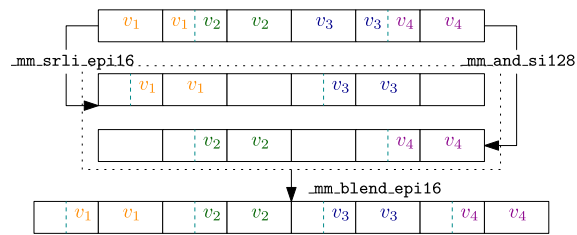}
  \caption{Simplified transformation of four packed 12-bit integers to 16-bit integers.
    Each block represents 8 bit.
    The split blocks contain 4 bit of two different entries.}
  \label{fig:example_12_to_16_bit}
\end{figure}

Therefore, our first objective is to transform the seven entries to use 16 bits during the comparison.
Of course, we cannot simply store the entries using 16 bits each, as this would increase the memory requirements significantly.
Instead, we have to unpack the 12-bit entries into consecutive 16-bit words, see \cref{fig:example_12_to_16_bit}.
To this end, we first consider the entries as 8-bit words.
Now, three 8-bit words contain two 12-bit entries: the first 8-bit word contains the upper part of the first entry, the second 8-bit word contains the lower part of the first entry and the upper part of the second entry, and the third 8-bit word contains the lower part of the second entry.
This pattern repeats for all entries.

We now can shift the words that have their upper part in a whole 8-bit word four bits to the right (\texttt{\_mm\_srli\_epi16}) and mask the lower 12 bit of the other entries (\texttt{\_mm\_set1\_epi16}) to obtain 16-bit words containing the entries as results.
Then, we can take one 16-bit word of each result alternately to obtain our final result, where each 12-bit entry is stored in consecutive 16-bit words.
Using this final result, we can compare  (\texttt{\_mm\_cmpgt\_epi16}) with the remaining rank minus one (because there does not exist a greater-or-equal comparison).
As the compare operation does not return the word, where the first match occurs, we have to transform the result to the position of the block.
The compare operation returns a word, where the result of the compare operation is marked by all-ones (true) or all-zeros (false).
Therefore, we can take the most significant bit of each 8-bit word (\texttt{\_mm\_movemask\_eop8}) and apply a simple popcount operation on the result, because we are only interested in the first result where the entry is greater or equal.
Now, we can continue the select query as before.

\subsection{Wide-Popcount: Faster Rank}
\label{sec:wide_popcount}
As mentioned before, using 16 bits to store an entry of the L2-index would allow us to directly use the SIMD instructions on the L2-index without transforming it first.
We do so in our final rank data structure that we call \emph{wide-popcount}.
Since we now have 16 bits available for each entry in the L2-index, we also have to make the L1-block bigger, because otherwise, the additional space required by the index would be too much.
Therefore, we do now let each L1-block cover 128 L2-blocks, or 65\,536 bits.
As before, we are only interested in the first 127 L2-blocks within each L1-block.
Thus, each entry in the L2-index is in \([0, 65\,024]\) and can be stored using 16 bits.
For the L1-index, we use 64-bit words, because then we do not need an L0-index.
This increases the required space of the data structure for rank queries to \(3.198\,\%\) additional space.
On the other hand, we have only two levels of indices instead of three.
We also do not interleave the L1- and L2-index, as there is no advantage because not all L2-entries can be loaded directly into the cache together with the L1-entry, due to their size and number.

Answering rank and select queries works the same as with flat-popcount (without an L0-index).
In \cref{sec:experimental_evaluation}, we will see that this approach works very well for rank queries, but it is not well suited for select queries.
This is mostly because we have to search through a lot of L2-entries during a select query.
Even when speeding up the search using a uniform binary search or SIMD instructions, answering select queries requires significantly more time than using flat-popcount.

\section{Experimental Evaluation}
\label{sec:experimental_evaluation} 

Our implementations are available at \url{https://github.com/pasta-toolbox/bit_vector} as open-source (GPLv3, header-only C++-library).
In addition to the source code of our implementations, we also provide scripts to easily reproduce all results presented in this paper (\url{https://github.com/pasta-toolbox/bit_vector_experiments}).
Our experiments were conducted on a workstation equipped with an AMD Ryzen 9 3950X (16 cores with frequencies up to 3.5\,GHz, 64\,KiB L1d and L1i cache and 512\,KiB L2 cache per core, and 4 times 16\,MiB L3 cache) and 64\,GiB DDR4 RAM running Ubuntu 20.04.2 LTS.
Since our experiments are sequential, only one core was used at a time.
We compiled the code using GCC 10.2 with flags \texttt{-03}, \texttt{-march=native}, and \texttt{-DNDEBUG} and created the makefile using CMake version 3.22.1.

In our experiments, we use two types of random inputs with different densities of ones in the bit vector (10\,\%, 50\,\%, and 90\,\% of all bits are ones).
For the first type of input, the ones are uniformly distributed.
This type of input should be the easiest one of the two.
The second type of input is an adversarial input similar to the one used by Vigna~\cite{Vigna2008BroadwordRankSelect}.
Assume that \(k\,\%\) of the bits in the bit vector should be ones.
Then, we set \(99\,\%\) of the ones in the last \(k\,\%\) of the bit vector and the remaining one percent in the first \(100-k\,\%\) of the bit vector.
Here, the first part of the bit vector is very sparse while the second part of the bit vector is very dense.
Overall, these two types of distribution are the extreme ends of distributions that can occur.
All data structures are tested on the same bit vectors and the same queries.
The reported running times are the average of three runs (each with a new bit vector and queries).
During each run, we constructed the data structure and then asked 100 million queries of each query type supported by the data structure.
We compare the following rank and select data structures:
\begin{itemize}
\item \emph{cs-poppy} is the space-efficient rank-based rank and select data structure described in \cref{sec:cs-poppy} by Zhou~et~al.~\cite{ZhouAK2013PopcountRankSelect},
\item \emph{cs-poppy-fs} is the same as cs-poppy but with the supposedly faster select algorithm used for the final 64-bit word by Pandey~et~al.~\cite{PandeyBJ2017FastSelect},
\item \emph{simple-select$_x$} is a position-based select data structure by Vigna~\cite{Vigna2008BroadwordRankSelect} that allows for tuning parameter \(x\) that determines the size of additional space the data structure is allowed to allocate,
\item \emph{simple-select$_h$} is a version of simple-select by Vigna~\cite{Vigna2008BroadwordRankSelect} that is highly tuned for bit vectors that contain the same amount of ones and zeros,
\item \emph{rank9select} is a rank and select data structure by Vigna~\cite{Vigna2008BroadwordRankSelect} that stores 9-bit values to answer rank queries and positions to answer select queries,
\item \emph{sdsl-mcl} Clark's select data structure \cite{Clark1997PhD} contained in the SDSL~\cite{GogBMP2014SDSL},
\item \emph{sdsl-v} is a simple rank data structure that requires 25\,\% additional memory and is contained in the SDSL, and
\item \emph{sdsl-v5} is a more space-efficient variant of the rank data structure above (only 6.25\,\% additional space) and also contained in the SDSL
\end{itemize}
with our implementations that we describe in \cref{sec:cs-poppy,sec:flat_popcount,sec:wide_popcount}:
\begin{itemize}
\item \emph{pasta-poppy} is our implementation of cs-poppy,
\item \emph{pasta-flat$_t$} is the rank-based rank and select data structure that we describe in \cref{sec:flat_popcount} with \(t\in\{\texttt{linear, binary, SIMD}\}\), and
\item \emph{pasta-wide} is the rank-data structure that we describe in \cref{sec:wide_popcount}.
\end{itemize}
Unfortunately, two competitors cs-poppy~\cite{ZhouAK2013PopcountRankSelect} and cs-poppy-fs~\cite{PandeyBJ2017FastSelect} were not able to compute the select queries on all inputs in a reasonable time (3 hours for all queries on a single bit vector).
We were not able to find the error causing this problem, but want to highlight that all queries asked were feasible queries.


\begin{table}[t]
  \centering
    \caption{Average additional space in percent used by all evaluated data structures on bit vectors of different sizes over the uniform and adversarial distribution.
    }
  \label{tab:space_requirements}
  \begin{tabular}{llrrrrrr}
    \toprule
    Name & \(n=\)& \(1\cdot 10^9\) & \(2\cdot 10^9\) & \(4\cdot 10^9\) & \(8\cdot 10^9\) & \(16\cdot 10^9\) & \(32\cdot 10^9\)\\
    \midrule
    cs-poppy &&  3.32 &  3.32 &  3.32 &       &       &       \\
    cs-poppy-fs &&  3.32 &  3.32 &  3.32 &       &       &       \\
    pasta-poppy &&  3.58 &  3.58 &  3.58 &  3.58 &  3.58 &  3.58 \\
    pasta-flat\(_{\texttt{SIMD}}\)&&  3.58 &  3.58 &  3.58 &  3.58 &  3.58 &  3.58 \\
    \midrule
    pasta-wide && 10.16 & 10.17 & 10.16 & 10.16 & 10.16 & 10.16 \\
    sdsl-v && 25.00 & 25.00 & 25.00 & 25.00 & 25.00 & 25.00 \\
    sdsl-v5 &&  6.25 &  6.25 &  6.25 &  6.25 &  6.25 &  6.25 \\
    \midrule
    sdsl-mcl && 18.51 & 18.52 & 18.53 & 18.54 & 18.55 & 18.56 \\
    simple-select\(_0\) &&  8.72 &  8.72 &  8.72 &  8.72 &  8.72 &  8.72 \\
    simple-select\(_1\) &&  9.88 &  9.88 &  9.88 &  9.88 &  9.88 &  9.88 \\
    simple-select\(_2\) && 12.21 & 12.20 & 12.20 & 12.20 & 12.20 & 12.20 \\
    simple-select\(_3\) && 16.85 & 16.85 & 16.84 & 16.84 & 16.84 & 16.84 \\
    simple-select\(_h\) && 15.62 & 15.63 & 15.63 & 15.63 & 15.63 & 15.63 \\
    rank9select && 56.25 & 56.25 & 56.25 & 56.25 & 56.25 & 56.25 \\
    \bottomrule
  \end{tabular}
\end{table}

We only include \emph{pasta-flat}\(_{\texttt{SIMD}}\) in the plots as it is overall the fastest variant.
For a comparison of the select query times of the three \emph{pasta-flat} query versions, see \cref{fig:select_query_times_pasta_only}.
The rank query times are identical, as the same data structure and query algorithm is used.
Note that we did not include the rank and select data structures that have already been shown to be slower (and to require more additional memory) than the ones included in the experimental evaluation, e.g., \emph{combined-sampling}~\cite{NavarroP2012CombinedSampling}, which is slower than cs-poppy and only works for bit vectors up to size \(2^{32}\) bits, \emph{BitRankF}~\cite{GonzalezGMN2005PracticalRankSelect}, which is slower than simple-select and also requires more space, and the data structures described by Kim et al.~\cite{KimNKP2005RankAndSelect}.

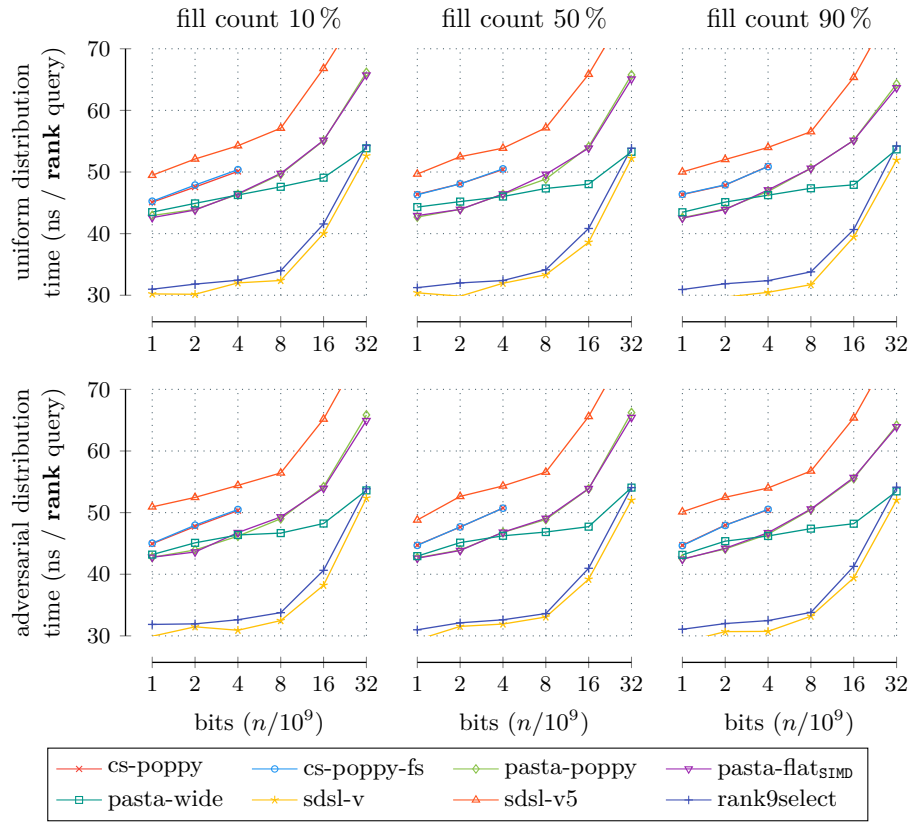
\begin{figure}[t]
\begin{center}
\begin{tikzpicture}
  \begin{axis}[
    plotRankQueryTimeAll,
    legend columns=4,
    legend to name={leg:rank_query_times},
    title={fill count 10\,\%},
    ylabel={\begin{tabular}{c}uniform distribution\\time (ns / \textbf{rank} query)\end{tabular}},
    xlabel near ticks,
    ylabel near ticks,
    xtick={0,1,2,3,4,5},
    xticklabels={1,2,4,8,16,32},
    ]

    \addplot coordinates { (0.0,45.1) (1.0,47.56) (2.0,50.1) };
    \addlegendentry{cs-poppy};
    \addplot coordinates { (0.0,45.26) (1.0,47.88) (2.0,50.38) };
    \addlegendentry{cs-poppy-fs};
    \addplot coordinates { (0.0,42.94) (1.0,43.92) (2.0,46.37) (3.0,49.63) (4.0,55.09) (5.0,66.16) };
    \addlegendentry{pasta-poppy};
    \addplot coordinates { (0.0,42.6) (1.0,43.83) (2.0,46.47) (3.0,49.77) (4.0,55.15) (5.0,65.72) };
    \addlegendentry{pasta-flat\(_{\texttt{SIMD}}\)};
    \addplot coordinates { (0.0,43.49) (1.0,44.91) (2.0,46.27) (3.0,47.6) (4.0,49.08) (5.0,53.85) };
    \addlegendentry{pasta-wide};
    \addplot coordinates { (0.0,30.22) (1.0,30.16) (2.0,32.01) (3.0,32.4) (4.0,40.02) (5.0,52.62) };
    \addlegendentry{sdsl-v};
    \addplot coordinates { (0.0,49.47) (1.0,52.11) (2.0,54.26) (3.0,57.11) (4.0,66.81) (5.0,77.8) };
    \addlegendentry{sdsl-v5};
    \addplot coordinates { (0.0,30.98) (1.0,31.81) (2.0,32.44) (3.0,33.98) (4.0,41.57) (5.0,54.3) };
    \addlegendentry{rank9select};

  \end{axis}
\end{tikzpicture}
\hspace{-.5cm}
\begin{tikzpicture}
  \begin{axis}[
    plotRankQueryTimeAll,
    title={fill count 50\,\%},
    y tick style={draw=none},
    yticklabel={\empty},
    y axis line style={draw=none},
    xlabel near ticks,
    xtick={0,1,2,3,4,5},
    xticklabels={1,2,4,8,16,32},
    ]

    \addplot coordinates { (0.0,46.38) (1.0,48.08) (2.0,50.39) };
    \addlegendentry{algo=efficient-poppy-rank-select};
    \addplot coordinates { (0.0,46.31) (1.0,48.07) (2.0,50.51) };
    \addlegendentry{algo=efficient-poppy-rank-select-improved};
    \addplot coordinates { (0.0,42.69) (1.0,43.94) (2.0,46.4) (3.0,48.85) (4.0,54.1) (5.0,65.76) };
    \addlegendentry{algo=pasta-popcount};
    \addplot coordinates { (0.0,42.93) (1.0,43.93) (2.0,46.43) (3.0,49.63) (4.0,53.87) (5.0,65.08) };
    \addlegendentry{algo=pasta-popcount-flat-one\_dont\_care-intrinsics};
    \addplot coordinates { (0.0,44.31) (1.0,45.2) (2.0,46.02) (3.0,47.33) (4.0,48.03) (5.0,53.28) };
    \addlegendentry{algo=pasta-popcount-wide-one\_dont\_care-linear\_search};
    \addplot coordinates { (0.0,30.4) (1.0,29.85) (2.0,31.95) (3.0,33.33) (4.0,38.6) (5.0,52.22) };
    \addlegendentry{algo=sdsl-rank-v};
    \addplot coordinates { (0.0,49.67) (1.0,52.48) (2.0,53.87) (3.0,57.17) (4.0,65.86) (5.0,77.57) };
    \addlegendentry{algo=sdsl-rank-v5};
    \addplot coordinates { (0.0,31.24) (1.0,32.0) (2.0,32.39) (3.0,34.14) (4.0,40.84) (5.0,53.86) };
    \addlegendentry{algo=sux-rank9select};
    \legend{};
  \end{axis}
\end{tikzpicture}
\hspace{-.5cm}
\begin{tikzpicture}
  \begin{axis}[
    plotRankQueryTimeAll,
    title={fill count 90\,\%},
    y tick style={draw=none},
    yticklabel={\empty},
    y axis line style={draw=none},
    xlabel near ticks,
    xtick={0,1,2,3,4,5},
    xticklabels={1,2,4,8,16,32},
    ]

    \addplot coordinates { (0.0,46.34) (1.0,47.87) (2.0,50.94) };
    \addlegendentry{algo=efficient-poppy-rank-select};
    \addplot coordinates { (0.0,46.39) (1.0,47.92) (2.0,50.88) };
    \addlegendentry{algo=efficient-poppy-rank-select-improved};
    \addplot coordinates { (0.0,42.59) (1.0,44.02) (2.0,46.84) (3.0,50.59) (4.0,55.14) (5.0,64.28) };
    \addlegendentry{algo=pasta-popcount};
    \addplot coordinates { (0.0,42.51) (1.0,43.91) (2.0,47.09) (3.0,50.6) (4.0,55.14) (5.0,63.66) };
    \addlegendentry{algo=pasta-popcount-flat-one\_dont\_care-intrinsics};
    \addplot coordinates { (0.0,43.47) (1.0,45.12) (2.0,46.24) (3.0,47.36) (4.0,47.92) (5.0,53.67) };
    \addlegendentry{algo=pasta-popcount-wide-one\_dont\_care-linear\_search};
    \addplot coordinates { (0.0,29.56) (1.0,29.7) (2.0,30.48) (3.0,31.72) (4.0,39.46) (5.0,51.93) };
    \addlegendentry{algo=sdsl-rank-v};
    \addplot coordinates { (0.0,50.01) (1.0,52.03) (2.0,53.99) (3.0,56.54) (4.0,65.35) (5.0,77.84) };
    \addlegendentry{algo=sdsl-rank-v5};
    \addplot coordinates { (0.0,30.93) (1.0,31.87) (2.0,32.37) (3.0,33.8) (4.0,40.67) (5.0,54.21) };
    \addlegendentry{algo=sux-rank9select};
    \legend{}
  \end{axis}
\end{tikzpicture}
\begin{tikzpicture}
  \begin{axis}[
    plotRankQueryTimeAll,
    xlabel={bits (\(n/10^9\))},
    ylabel={\begin{tabular}{c}adversarial distribution\\time (ns / \textbf{rank} query)\end{tabular}},
    xlabel near ticks,
    ylabel near ticks,
    xtick={0,1,2,3,4,5},
    xticklabels={1,2,4,8,16,32},
    ]

    \addplot coordinates { (0.0,44.95) (1.0,47.75) (2.0,50.33) };
    \addlegendentry{algo=efficient-poppy-rank-select};
    \addplot coordinates { (0.0,45.03) (1.0,47.96) (2.0,50.5) };
    \addlegendentry{algo=efficient-poppy-rank-select-improved};
    \addplot coordinates { (0.0,42.75) (1.0,43.94) (2.0,46.18) (3.0,49.01) (4.0,54.19) (5.0,65.85) };
    \addlegendentry{algo=pasta-popcount};
    \addplot coordinates { (0.0,42.76) (1.0,43.61) (2.0,46.76) (3.0,49.3) (4.0,53.94) (5.0,64.93) };
    \addlegendentry{algo=pasta-popcount-flat-one\_dont\_care-intrinsics};
    \addplot coordinates { (0.0,43.19) (1.0,45.1) (2.0,46.37) (3.0,46.68) (4.0,48.23) (5.0,53.61) };
    \addlegendentry{algo=pasta-popcount-wide-one\_dont\_care-linear\_search};
    \addplot coordinates { (0.0,29.94) (1.0,31.47) (2.0,30.91) (3.0,32.48) (4.0,38.21) (5.0,52.36) };
    \addlegendentry{algo=sdsl-rank-v};
    \addplot coordinates { (0.0,50.93) (1.0,52.47) (2.0,54.42) (3.0,56.44) (4.0,65.18) (5.0,77.6) };
    \addlegendentry{algo=sdsl-rank-v5};
    \addplot coordinates { (0.0,31.86) (1.0,31.95) (2.0,32.62) (3.0,33.77) (4.0,40.64) (5.0,53.87) };
    \addlegendentry{algo=sux-rank9select};
    \legend{};
  \end{axis}
\end{tikzpicture}
\hspace{-.5cm}
\begin{tikzpicture}
  \begin{axis}[
    plotRankQueryTimeAll,
    xlabel={bits (\(n/10^9\))},
    y tick style={draw=none},
    yticklabel={\empty},
    y axis line style={draw=none},
    xlabel near ticks,
    xtick={0,1,2,3,4,5},
    xticklabels={1,2,4,8,16,32},
    ]

    \addplot coordinates { (0.0,44.74) (1.0,47.65) (2.0,50.75) };
    \addlegendentry{algo=efficient-poppy-rank-select};
    \addplot coordinates { (0.0,44.69) (1.0,47.68) (2.0,50.72) };
    \addlegendentry{algo=efficient-poppy-rank-select-improved};
    \addplot coordinates { (0.0,42.67) (1.0,43.93) (2.0,46.79) (3.0,48.88) (4.0,53.85) (5.0,66.23) };
    \addlegendentry{algo=pasta-popcount};
    \addplot coordinates { (0.0,42.6) (1.0,43.85) (2.0,46.82) (3.0,49.07) (4.0,53.9) (5.0,65.44) };
    \addlegendentry{algo=pasta-popcount-flat-one\_dont\_care-intrinsics};
    \addplot coordinates { (0.0,42.94) (1.0,45.13) (2.0,46.24) (3.0,46.85) (4.0,47.72) (5.0,54.06) };
    \addlegendentry{algo=pasta-popcount-wide-one\_dont\_care-linear\_search};
    \addplot coordinates { (0.0,29.35) (1.0,31.54) (2.0,31.92) (3.0,33.06) (4.0,39.19) (5.0,52.02) };
    \addlegendentry{algo=sdsl-rank-v};
    \addplot coordinates { (0.0,48.8) (1.0,52.63) (2.0,54.33) (3.0,56.57) (4.0,65.6) (5.0,77.96) };
    \addlegendentry{algo=sdsl-rank-v5};
    \addplot coordinates { (0.0,30.99) (1.0,32.12) (2.0,32.61) (3.0,33.62) (4.0,40.97) (5.0,54.06) };
    \addlegendentry{algo=sux-rank9select};
    \legend{};
  \end{axis}
\end{tikzpicture}
\hspace{-.5cm}
\begin{tikzpicture}
  \begin{axis}[
    plotRankQueryTimeAll,
    xlabel={bits (\(n/10^9\))},
    y tick style={draw=none},
    yticklabel={\empty},
    y axis line style={draw=none},
    xlabel near ticks,
    xtick={0,1,2,3,4,5},
    xticklabels={1,2,4,8,16,32},
    ]

    \addplot coordinates { (0.0,44.71) (1.0,47.98) (2.0,50.45) };
    \addlegendentry{efficient-poppy-rank-select};
    \addplot coordinates { (0.0,44.65) (1.0,47.93) (2.0,50.53) };
    \addlegendentry{efficient-poppy-rank-select-improved};
    \addplot coordinates { (0.0,42.49) (1.0,44.1) (2.0,46.5) (3.0,50.48) (4.0,55.53) (5.0,64.17) };
    \addlegendentry{pasta-popcount};
    \addplot coordinates { (0.0,42.43) (1.0,44.21) (2.0,46.74) (3.0,50.58) (4.0,55.7) (5.0,63.92) };
    \addlegendentry{pasta-popcount-flat-one\_dont\_care-intrinsics};
    \addplot coordinates { (0.0,43.15) (1.0,45.37) (2.0,46.21) (3.0,47.4) (4.0,48.2) (5.0,53.45) };
    \addlegendentry{pasta-popcount-wide-one\_dont\_care-linear\_search};
    \addplot coordinates { (0.0,28.95) (1.0,30.69) (2.0,30.73) (3.0,33.19) (4.0,39.41) (5.0,52.03) };
    \addlegendentry{sdsl-rank-v};
    \addplot coordinates { (0.0,50.13) (1.0,52.49) (2.0,54.0) (3.0,56.75) (4.0,65.36) (5.0,77.71) };
    \addlegendentry{sdsl-rank-v5};
    \addplot coordinates { (0.0,31.07) (1.0,32.0) (2.0,32.48) (3.0,33.81) (4.0,41.28) (5.0,54.13) };
    \addlegendentry{sux-rank9select};
    \legend{}
  \end{axis}
\end{tikzpicture}
\ref{leg:rank_query_times}
\caption{Average rank query time in nanoseconds on all tested inputs.}
\label{fig:rank_query_times}
\end{center}
\end{figure}

\paragraph*{Space Requirements and Select\(_0\) \& Select\(_1\) Queries.}
First, we discuss the space requirements of all data structures evaluated in this paper.
For an overview, see \cref{tab:space_requirements}.
We measured the additional space by overwriting \emph{malloc} to see all allocations on the heap of the different data structures.
This is also the reason why it looks like cs-poppy and cs-poppy-fs require less space than pasta-poppy and pasta-flat, because the former allocates memory on the stack to store the samples for the select queries (which we did not modify to not change the results of the running time experiments).
This makes our new data structures, cs-poppy, and cs-poppy-fs the most space-efficient rank and select data structures, requiring roughly half the space that the smallest rank- and select-only data structures (sdsl-v5 and simple-select\(_0\)) require.

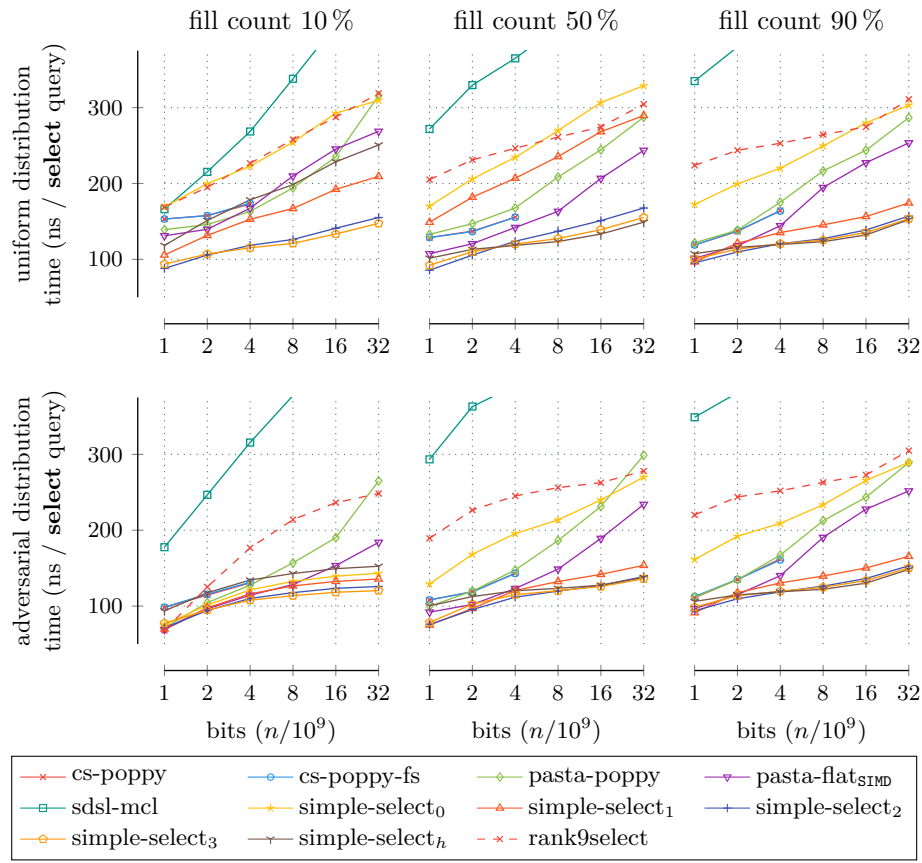
\begin{figure}[t]
\begin{center}
\begin{tikzpicture}
  \begin{axis}[
    plotSelectQueryTimeAll,
    legend columns=4,
    legend to name={leg:select_query_times},
    title={fill count 10\,\%},
    ylabel={\begin{tabular}{c}uniform distribution\\time (ns / \textbf{select} query)\end{tabular}},
    xlabel near ticks,
    ylabel near ticks,
    xtick={0,1,2,3,4,5},
    xticklabels={1,2,4,8,16,32},
    ]

    \addplot coordinates { (0.0,153.14) (1.0,157.63) (2.0,172.7) };
    \addlegendentry{cs-poppy};
    \addplot coordinates { (0.0,153.16) (1.0,157.66) (2.0,174.21) };
    \addlegendentry{cs-poppy-fs};
    \addplot coordinates { (0.0,139.15) (1.0,145.89) (2.0,163.17) (3.0,194.18) (4.0,235.19) (5.0,315.91) };
    \addlegendentry{pasta-poppy};
    \addplot coordinates { (0.0,131.01) (1.0,139.59) (2.0,167.48) (3.0,209.81) (4.0,245.45) (5.0,268.77) };
    \addlegendentry{pasta-flat\(_\texttt{SIMD}\)};
    \addplot coordinates { (0.0,166.1) (1.0,215.36) (2.0,268.58) (3.0,338.03) (4.0,409.15) (5.0,446.97) };
    \addlegendentry{sdsl-mcl};
    \addplot coordinates { (0.0,169.06) (1.0,199.74) (2.0,222.82) (3.0,254.36) (4.0,292.72) (5.0,309.72) };
    \addlegendentry{simple-select\(_0\)};
    \addplot coordinates { (0.0,105.77) (1.0,131.56) (2.0,152.83) (3.0,166.99) (4.0,192.32) (5.0,209.36) };
    \addlegendentry{simple-select\(_1\)};
    \addplot coordinates { (0.0,87.99) (1.0,105.61) (2.0,118.45) (3.0,126.18) (4.0,140.99) (5.0,155.09) };
    \addlegendentry{simple-select\(_2\)};
    \addplot coordinates { (0.0,93.72) (1.0,106.92) (2.0,115.27) (3.0,121.06) (4.0,133.47) (5.0,147.39) };
    \addlegendentry{simple-select\(_3\)};
    \addplot coordinates { (0.0,118.34) (1.0,151.3) (2.0,178.49) (3.0,198.51) (4.0,228.6) (5.0,250.77) };
    \addlegendentry{simple-select\(_h\)};
    \addplot coordinates { (0.0,169.03) (1.0,194.94) (2.0,226.61) (3.0,257.92) (4.0,287.59) (5.0,318.8) };
    \addlegendentry{rank9select};

  \end{axis}
\end{tikzpicture}
\hspace{-.5cm}
\begin{tikzpicture}
  \begin{axis}[
    plotSelectQueryTimeAll,
    title={fill count 50\,\%},
    y tick style={draw=none},
    yticklabel={\empty},
    y axis line style={draw=none},
    xlabel near ticks,
    xtick={0,1,2,3,4,5},
    xticklabels={1,2,4,8,16,32},
    ]

    \addplot coordinates { (0.0,128.66) (1.0,137.18) (2.0,155.56) };
    \addlegendentry{algo=efficient-poppy-rank-select};
    \addplot coordinates { (0.0,128.76) (1.0,136.57) (2.0,155.58) };
    \addlegendentry{algo=efficient-poppy-rank-select-improved};
    \addplot coordinates { (0.0,133.0) (1.0,146.96) (2.0,167.73) (3.0,208.46) (4.0,244.45) (5.0,287.39) };
    \addlegendentry{algo=pasta-popcount};
    \addplot coordinates { (0.0,107.38) (1.0,120.58) (2.0,141.96) (3.0,163.26) (4.0,206.73) (5.0,243.72) };
    \addlegendentry{algo=pasta-popcount-flat-one\_dont\_care-intrinsics};
    \addplot coordinates { (0.0,271.85) (1.0,329.67) (2.0,365.0) (3.0,406.64) (4.0,444.99) (5.0,478.83) };
    \addlegendentry{algo=sdsl-select-mcl};
    \addplot coordinates { (0.0,170.15) (1.0,205.98) (2.0,234.37) (3.0,269.77) (4.0,306.67) (5.0,328.9) };
    \addlegendentry{algo=sux-SimpleSelect-0};
    \addplot coordinates { (0.0,148.92) (1.0,181.9) (2.0,206.97) (3.0,235.45) (4.0,268.24) (5.0,289.66) };
    \addlegendentry{algo=sux-SimpleSelect-1};
    \addplot coordinates { (0.0,85.68) (1.0,105.52) (2.0,123.89) (3.0,136.98) (4.0,151.06) (5.0,167.66) };
    \addlegendentry{algo=sux-SimpleSelect-2};
    \addplot coordinates { (0.0,92.22) (1.0,110.15) (2.0,120.13) (3.0,127.44) (4.0,139.16) (5.0,155.24) };
    \addlegendentry{algo=sux-SimpleSelect-3};
    \addplot coordinates { (0.0,101.5) (1.0,113.17) (2.0,118.36) (3.0,123.46) (4.0,133.35) (5.0,148.96) };
    \addlegendentry{algo=sux-SimpleSelectHalf};
    \addplot coordinates { (0.0,205.17) (1.0,231.18) (2.0,246.41) (3.0,261.23) (4.0,274.54) (5.0,304.51) };
    \addlegendentry{algo=sux-rank9select};
    \legend{};
  \end{axis}
\end{tikzpicture}
\hspace{-.5cm}
\begin{tikzpicture}
  \begin{axis}[
    plotSelectQueryTimeAll,
    title={fill count 90\,\%},
    y tick style={draw=none},
    yticklabel={\empty},
    y axis line style={draw=none},
    xlabel near ticks,
    xtick={0,1,2,3,4,5},
    xticklabels={1,2,4,8,16,32},
    ]

    \addplot coordinates { (0.0,119.05) (1.0,136.82) (2.0,164.32) };
    \addlegendentry{algo=efficient-poppy-rank-select};
    \addplot coordinates { (0.0,118.89) (1.0,137.64) (2.0,163.63) };
    \addlegendentry{algo=efficient-poppy-rank-select-improved};
    \addplot coordinates { (0.0,121.73) (1.0,138.53) (2.0,175.26) (3.0,216.56) (4.0,243.83) (5.0,286.83) };
    \addlegendentry{algo=pasta-popcount};
    \addplot coordinates { (0.0,101.26) (1.0,117.81) (2.0,144.27) (3.0,194.66) (4.0,227.34) (5.0,253.52) };
    \addlegendentry{algo=pasta-popcount-flat-one\_dont\_care-intrinsics};
    \addplot coordinates { (0.0,335.04) (1.0,379.97) (2.0,406.45) (3.0,431.08) (4.0,470.23) (5.0,521.68) };
    \addlegendentry{algo=sdsl-select-mcl};
    \addplot coordinates { (0.0,172.34) (1.0,199.41) (2.0,220.14) (3.0,249.48) (4.0,279.79) (5.0,303.37) };
    \addlegendentry{algo=sux-SimpleSelect-0};
    \addplot coordinates { (0.0,96.59) (1.0,121.28) (2.0,135.23) (3.0,145.44) (4.0,156.47) (5.0,174.28) };
    \addlegendentry{algo=sux-SimpleSelect-1};
    \addplot coordinates { (0.0,95.44) (1.0,109.7) (2.0,120.56) (3.0,127.24) (4.0,138.67) (5.0,157.0) };
    \addlegendentry{algo=sux-SimpleSelect-2};
    \addplot coordinates { (0.0,100.67) (1.0,113.4) (2.0,120.69) (3.0,125.39) (4.0,135.38) (5.0,153.5) };
    \addlegendentry{algo=sux-SimpleSelect-3};
    \addplot coordinates { (0.0,107.38) (1.0,115.7) (2.0,119.6) (3.0,123.25) (4.0,132.17) (5.0,152.33) };
    \addlegendentry{algo=sux-SimpleSelectHalf};
    \addplot coordinates { (0.0,223.83) (1.0,243.74) (2.0,253.0) (3.0,264.44) (4.0,274.6) (5.0,311.05) };
    \addlegendentry{algo=sux-rank9select};
    \legend{}
  \end{axis}
\end{tikzpicture}
\begin{tikzpicture}
  \begin{axis}[
    plotSelectQueryTimeAll,
    xlabel={bits (\(n/10^9\))},
    ylabel={\begin{tabular}{c}adversarial distribution\\time (ns / \textbf{select} query)\end{tabular}},
    xlabel near ticks,
    ylabel near ticks,
    xtick={0,1,2,3,4,5},
    xticklabels={1,2,4,8,16,32},
    ]

    \addplot coordinates { (0.0,98.18) (1.0,114.46) (2.0,130.55) };
    \addlegendentry{algo=efficient-poppy-rank-select};
    \addplot coordinates { (0.0,98.63) (1.0,115.67) (2.0,131.37) };
    \addlegendentry{algo=efficient-poppy-rank-select-improved};
    \addplot coordinates { (0.0,71.36) (1.0,104.54) (2.0,127.54) (3.0,156.93) (4.0,190.17) (5.0,265.06) };
    \addlegendentry{algo=pasta-popcount};
    \addplot coordinates { (0.0,69.07) (1.0,96.49) (2.0,114.29) (3.0,128.45) (4.0,153.28) (5.0,184.03) };
    \addlegendentry{algo=pasta-popcount-flat-one\_dont\_care-intrinsics};
    \addplot coordinates { (0.0,177.64) (1.0,246.64) (2.0,315.65) (3.0,377.5) (4.0,404.54) (5.0,425.76) };
    \addlegendentry{algo=sdsl-select-mcl};
    \addplot coordinates { (0.0,75.35) (1.0,101.36) (2.0,121.39) (3.0,132.68) (4.0,139.39) (5.0,143.27) };
    \addlegendentry{algo=sux-SimpleSelect-0};
    \addplot coordinates { (0.0,69.86) (1.0,97.79) (2.0,116.1) (3.0,126.62) (4.0,132.51) (5.0,135.68) };
    \addlegendentry{algo=sux-SimpleSelect-1};
    \addplot coordinates { (0.0,71.39) (1.0,94.21) (2.0,110.12) (3.0,117.72) (4.0,123.43) (5.0,125.84) };
    \addlegendentry{algo=sux-SimpleSelect-2};
    \addplot coordinates { (0.0,77.38) (1.0,94.79) (2.0,107.83) (3.0,113.74) (4.0,118.19) (5.0,120.4) };
    \addlegendentry{algo=sux-SimpleSelect-3};
    \addplot coordinates { (0.0,93.69) (1.0,117.54) (2.0,134.33) (3.0,142.91) (4.0,149.33) (5.0,152.52) };
    \addlegendentry{algo=sux-SimpleSelectHalf};
    \addplot coordinates { (0.0,67.95) (1.0,125.17) (2.0,176.75) (3.0,213.98) (4.0,236.38) (5.0,248.4) };
    \addlegendentry{algo=sux-rank9select};
    \legend{};
  \end{axis}
\end{tikzpicture}
\hspace{-.5cm}
\begin{tikzpicture}
  \begin{axis}[
    plotSelectQueryTimeAll,
    xlabel={bits (\(n/10^9\))},
    y tick style={draw=none},
    yticklabel={\empty},
    y axis line style={draw=none},
    xlabel near ticks,
    xtick={0,1,2,3,4,5},
    xticklabels={1,2,4,8,16,32},
    ]

    \addplot coordinates { (0.0,108.11) (1.0,118.98) (2.0,143.2) };
    \addlegendentry{algo=efficient-poppy-rank-select};
    \addplot coordinates { (0.0,108.16) (1.0,118.67) (2.0,143.04) };
    \addlegendentry{algo=efficient-poppy-rank-select-improved};
    \addplot coordinates { (0.0,99.7) (1.0,119.75) (2.0,147.52) (3.0,186.35) (4.0,231.43) (5.0,298.94) };
    \addlegendentry{algo=pasta-popcount};
    \addplot coordinates { (0.0,91.99) (1.0,102.09) (2.0,123.13) (3.0,148.73) (4.0,189.19) (5.0,234.07) };
    \addlegendentry{algo=pasta-popcount-flat-one\_dont\_care-intrinsics};
    \addplot coordinates { (0.0,293.46) (1.0,363.11) (2.0,390.55) (3.0,415.43) (4.0,443.37) (5.0,483.76) };
    \addlegendentry{algo=sdsl-select-mcl};
    \addplot coordinates { (0.0,129.14) (1.0,168.24) (2.0,195.55) (3.0,213.63) (4.0,239.93) (5.0,270.06) };
    \addlegendentry{algo=sux-SimpleSelect-0};
    \addplot coordinates { (0.0,75.05) (1.0,96.81) (2.0,120.63) (3.0,132.33) (4.0,141.83) (5.0,153.72) };
    \addlegendentry{algo=sux-SimpleSelect-1};
    \addplot coordinates { (0.0,75.63) (1.0,95.39) (2.0,111.74) (3.0,119.73) (4.0,127.29) (5.0,138.89) };
    \addlegendentry{algo=sux-SimpleSelect-2};
    \addplot coordinates { (0.0,78.33) (1.0,102.63) (2.0,115.08) (3.0,120.27) (4.0,125.85) (5.0,135.93) };
    \addlegendentry{algo=sux-SimpleSelect-3};
    \addplot coordinates { (0.0,100.33) (1.0,112.55) (2.0,120.13) (3.0,123.41) (4.0,127.55) (5.0,137.81) };
    \addlegendentry{algo=sux-SimpleSelectHalf};
    \addplot coordinates { (0.0,189.31) (1.0,226.53) (2.0,245.22) (3.0,256.21) (4.0,262.72) (5.0,278.07) };
    \addlegendentry{algo=sux-rank9select};
    \legend{};
  \end{axis}
\end{tikzpicture}
\hspace{-.5cm}
\begin{tikzpicture}
  \begin{axis}[
    plotSelectQueryTimeAll,
    xlabel={bits (\(n/10^9\))},
    y tick style={draw=none},
    yticklabel={\empty},
    y axis line style={draw=none},
    xlabel near ticks,
    xtick={0,1,2,3,4,5},
    xticklabels={1,2,4,8,16,32},
    ]

    \addplot coordinates { (0.0,111.04) (1.0,134.93) (2.0,162.35) };
    \addlegendentry{efficient-poppy-rank-select};
    \addplot coordinates { (0.0,112.68) (1.0,135.26) (2.0,160.73) };
    \addlegendentry{efficient-poppy-rank-select-improved};
    \addplot coordinates { (0.0,111.69) (1.0,134.67) (2.0,167.16) (3.0,212.58) (4.0,243.41) (5.0,289.18) };
    \addlegendentry{pasta-popcount};
    \addplot coordinates { (0.0,96.58) (1.0,115.05) (2.0,140.03) (3.0,190.56) (4.0,227.74) (5.0,251.77) };
    \addlegendentry{pasta-popcount-flat-one\_dont\_care-intrinsics};
    \addplot coordinates { (0.0,348.99) (1.0,381.21) (2.0,404.71) (3.0,433.1) (4.0,471.61) (5.0,522.83) };
    \addlegendentry{sdsl-select-mcl};
    \addplot coordinates { (0.0,161.55) (1.0,192.03) (2.0,209.07) (3.0,233.51) (4.0,265.63) (5.0,289.44) };
    \addlegendentry{sux-SimpleSelect-0};
    \addplot coordinates { (0.0,91.32) (1.0,117.59) (2.0,130.54) (3.0,139.65) (4.0,150.26) (5.0,165.6) };
    \addlegendentry{sux-SimpleSelect-1};
    \addplot coordinates { (0.0,93.81) (1.0,109.63) (2.0,118.79) (3.0,126.24) (4.0,136.34) (5.0,153.14) };
    \addlegendentry{sux-SimpleSelect-2};
    \addplot coordinates { (0.0,98.6) (1.0,113.77) (2.0,119.61) (3.0,124.83) (4.0,133.6) (5.0,150.33) };
    \addlegendentry{sux-SimpleSelect-3};
    \addplot coordinates { (0.0,106.18) (1.0,114.65) (2.0,118.81) (3.0,122.18) (4.0,130.22) (5.0,148.31) };
    \addlegendentry{sux-SimpleSelectHalf};
    \addplot coordinates { (0.0,220.23) (1.0,243.89) (2.0,252.01) (3.0,263.14) (4.0,272.9) (5.0,304.94) };
    \addlegendentry{sux-rank9select};
    \legend{}
  \end{axis}
\end{tikzpicture}

\ref{leg:select_query_times}
\caption{Average select query time in nanoseconds on all tested inputs.}
\label{fig:select_query_times}
\end{center}
\end{figure}

\paragraph*{Rank Queries.}
Let us take a look at all data structures that support rank queries.
In \cref{fig:rank_query_times}, we report the average query time on all tested inputs.
Here, we can see that sdsl-v and rank9select provide the fastest query times.
For large inputs, pasta-wide has query times similar to sdsl-v and rank9select.
All these data structures can only answer rank queries and require more space than our new data structures (1.75--6.98 times as much).
Both pasta-poppy and pasta-flat have similar query times and get slower for larger inputs.
Nevertheless, they are roughly 8\,\% faster than cs-poppy and cs-poppy-fs.

\paragraph*{Select Queries.}
We report select query times in \cref{fig:select_query_times}.
Here, we can see that sdsl-mcl, rank9select, and simple-select\(_0\) are among the slowest approaches.
Depending on the input, either simple-select\(_1\) or simple-select\(_2\) is always faster than our new data structures.
All other evaluated select data structures are somewhere between pasta-flat and simple-select\(_2\) when it comes to select query times.
This comes without surprise, simple-select are highly tuned select-only data structures that also use at least 2.43 times as much memory as pasta-flat.
However, pasta-flat\(_{\texttt{SIMD}}\) is 16.5\,\% faster than cs-poppy and cs-poppy-fs, making it the fastest and most space-efficient uncompressed rank \emph{and} select data structure.
  
\section{Conclusion}
With pasta-flat, we present a space-efficient rank and select data structure that is fast in practice.
It requires the same space as the  previously most space-efficient rank and select data structure cs-poppy and is between 8\,\% (rank) and 16.5\,\% (select) faster than cs-poppy.
While there exist faster rank- and select-only data structures, they require significantly more memory and cannot easily answer both \(select_0\) and \(select_1\) queries, a necessity for many applications, e.g., wavelet trees~\cite{FerraginaM2000FMIndex} or succinct tree representations~\cite{Jacobson1989LOUDS}.
Pasta-flat can answer both (with a slowdown of up to 1.5 for one of the queries) without requiring additional memory.

\subsubsection*{Acknowledgements.}
This project has received funding from the European Research Council (ERC) under the European Union’s Horizon 2020 research and innovation programme (grant agreement No. 882500).

\begin{center}
  \includegraphics[width=4.5cm]{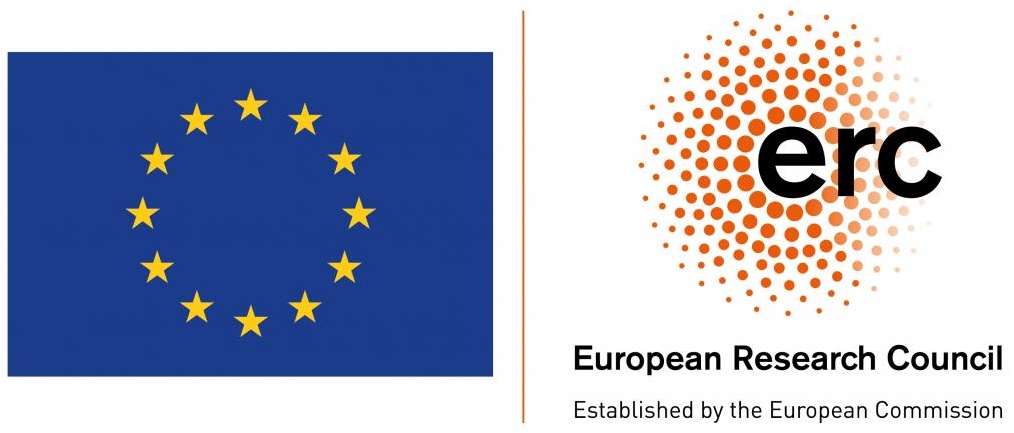}
\end{center}

%
%
%
\bibliography{lit}

\clearpage

\appendix

\section{Additional Experimental Results}

\subsection{Comparison of Our Implementations Only}
\begin{figure}[h!]
\begin{center}
\begin{tikzpicture}
  \begin{axis}[
    plotSelectQueryTimeAll,
    legend columns=4,
    legend to name={leg:select_query_times_pasta_only},
    title={fill count 10\,\%},
    ylabel={\begin{tabular}{c}uniform distribution\\time (ns / \textbf{select} query)\end{tabular}},
    xlabel near ticks,
    ylabel near ticks,
    xtick={0,1,2,3,4,5},
    xticklabels={1,2,4,8,16,32},
    ymin=90,
    ymax=325,
    height=37.5mm,
    ]

    \addplot coordinates { (0.0,139.15) (1.0,145.89) (2.0,163.17) (3.0,194.18) (4.0,235.19) (5.0,315.91) };
    \addlegendentry{pasta-poppy};
    \addplot coordinates { (0.0,128.07) (1.0,136.04) (2.0,169.78) (3.0,211.63) (4.0,247.18) (5.0,270.82) };
    \addlegendentry{pasta-flat\(_\texttt{binary}\)};
    \addplot coordinates { (0.0,131.01) (1.0,139.59) (2.0,167.48) (3.0,209.81) (4.0,245.45) (5.0,268.77) };
    \addlegendentry{pasta-flat\(_\texttt{SIMD}\)};
    \addplot coordinates { (0.0,133.73) (1.0,141.63) (2.0,171.56) (3.0,212.46) (4.0,249.75) (5.0,272.53) };
    \addlegendentry{pasta-flat\(_\texttt{linear}\)};

  \end{axis}
\end{tikzpicture}
\hspace{-.5cm}
\begin{tikzpicture}
  \begin{axis}[
    plotSelectQueryTimeAll,
    title={fill count 50\,\%},
    y tick style={draw=none},
    yticklabel={\empty},
    y axis line style={draw=none},
    xlabel near ticks,
    xtick={0,1,2,3,4,5},
    xticklabels={1,2,4,8,16,32},
    ymin=90,
    ymax=325,
    height=37.5mm,
    ]

    \addplot coordinates { (0.0,133.0) (1.0,146.96) (2.0,167.73) (3.0,208.46) (4.0,244.45) (5.0,287.39) };
    \addlegendentry{algo=pasta-popcount};
    \addplot coordinates { (0.0,118.43) (1.0,138.44) (2.0,162.47) (3.0,198.35) (4.0,230.7) (5.0,247.8) };
    \addlegendentry{algo=pasta-popcount-flat-one\_dont\_care-binary\_search};
    \addplot coordinates { (0.0,107.38) (1.0,120.58) (2.0,141.96) (3.0,163.26) (4.0,206.73) (5.0,243.72) };
    \addlegendentry{algo=pasta-popcount-flat-one\_dont\_care-intrinsics};
    \addplot coordinates { (0.0,116.3) (1.0,132.32) (2.0,153.41) (3.0,190.07) (4.0,229.61) (5.0,249.5) };
    \addlegendentry{algo=pasta-popcount-flat-one\_dont\_care-linear\_search};
    \legend{};
  \end{axis}
\end{tikzpicture}
\hspace{-.5cm}
\begin{tikzpicture}
  \begin{axis}[
    plotSelectQueryTimeAll,
    title={fill count 90\,\%},
    y tick style={draw=none},
    yticklabel={\empty},
    y axis line style={draw=none},
    xlabel near ticks,
    xtick={0,1,2,3,4,5},
    xticklabels={1,2,4,8,16,32},
    ymin=90,
    ymax=325,
    height=37.5mm,
    ]

    \addplot coordinates { (0.0,121.73) (1.0,138.53) (2.0,175.26) (3.0,216.56) (4.0,243.83) (5.0,286.83) };
    \addlegendentry{algo=pasta-popcount};
    \addplot coordinates { (0.0,122.65) (1.0,146.65) (2.0,187.93) (3.0,224.59) (4.0,241.34) (5.0,254.98) };
    \addlegendentry{algo=pasta-popcount-flat-one\_dont\_care-binary\_search};
    \addplot coordinates { (0.0,101.26) (1.0,117.81) (2.0,144.27) (3.0,194.66) (4.0,227.34) (5.0,253.52) };
    \addlegendentry{algo=pasta-popcount-flat-one\_dont\_care-intrinsics};
    \addplot coordinates { (0.0,112.8) (1.0,136.82) (2.0,180.63) (3.0,227.15) (4.0,246.44) (5.0,260.87) };
    \addlegendentry{algo=pasta-popcount-flat-one\_dont\_care-linear\_search};
    \legend{}
  \end{axis}
\end{tikzpicture}
\begin{tikzpicture}
  \begin{axis}[
    plotSelectQueryTimeAll,
    xlabel={bits (\(n/10^9\))},
    ylabel={\begin{tabular}{c}adversarial distribution\\time (ns / \textbf{select} query)\end{tabular}},
    xlabel near ticks,
    ylabel near ticks,
    xtick={0,1,2,3,4,5},
    xticklabels={1,2,4,8,16,32},
    ymin=90,
    ymax=325,
    height=37.5mm,
    ]

    \addplot coordinates { (0.0,71.36) (1.0,104.54) (2.0,127.54) (3.0,156.93) (4.0,190.17) (5.0,265.06) };
    \addlegendentry{algo=pasta-popcount};
    \addplot coordinates { (0.0,68.29) (1.0,99.24) (2.0,117.18) (3.0,136.82) (4.0,176.24) (5.0,217.16) };
    \addlegendentry{algo=pasta-popcount-flat-one\_dont\_care-binary\_search};
    \addplot coordinates { (0.0,69.07) (1.0,96.49) (2.0,114.29) (3.0,128.45) (4.0,153.28) (5.0,184.03) };
    \addlegendentry{algo=pasta-popcount-flat-one\_dont\_care-intrinsics};
    \addplot coordinates { (0.0,68.84) (1.0,101.02) (2.0,118.01) (3.0,133.44) (4.0,167.45) (5.0,211.19) };
    \addlegendentry{algo=pasta-popcount-flat-one\_dont\_care-linear\_search};
    \legend{};
  \end{axis}
\end{tikzpicture}
\hspace{-.5cm}
\begin{tikzpicture}
  \begin{axis}[
    plotSelectQueryTimeAll,
    xlabel={bits (\(n/10^9\))},
    y tick style={draw=none},
    yticklabel={\empty},
    y axis line style={draw=none},
    xlabel near ticks,
    xtick={0,1,2,3,4,5},
    xticklabels={1,2,4,8,16,32},
    ymin=90,
    ymax=325,
    height=37.5mm,
    ]

    \addplot coordinates { (0.0,99.7) (1.0,119.75) (2.0,147.52) (3.0,186.35) (4.0,231.43) (5.0,298.94) };
    \addlegendentry{algo=pasta-popcount};
    \addplot coordinates { (0.0,97.99) (1.0,124.08) (2.0,154.77) (3.0,193.15) (4.0,225.05) (5.0,246.78) };
    \addlegendentry{algo=pasta-popcount-flat-one\_dont\_care-binary\_search};
    \addplot coordinates { (0.0,91.99) (1.0,102.09) (2.0,123.13) (3.0,148.73) (4.0,189.19) (5.0,234.07) };
    \addlegendentry{algo=pasta-popcount-flat-one\_dont\_care-intrinsics};
    \addplot coordinates { (0.0,96.93) (1.0,114.64) (2.0,143.34) (3.0,186.95) (4.0,226.63) (5.0,252.3) };
    \addlegendentry{algo=pasta-popcount-flat-one\_dont\_care-linear\_search};
    \legend{};
  \end{axis}
\end{tikzpicture}
\hspace{-.5cm}
\begin{tikzpicture}
  \begin{axis}[
    plotSelectQueryTimeAll,
    xlabel={bits (\(n/10^9\))},
    y tick style={draw=none},
    yticklabel={\empty},
    y axis line style={draw=none},
    xlabel near ticks,
    xtick={0,1,2,3,4,5},
    xticklabels={1,2,4,8,16,32},
    ymin=90,
    ymax=325,
    height=37.5mm,
    ]

    \addplot coordinates { (0.0,111.69) (1.0,134.67) (2.0,167.16) (3.0,212.58) (4.0,243.41) (5.0,289.18) };
    \addlegendentry{pasta-poppy};
    \addplot coordinates { (0.0,113.77) (1.0,145.43) (2.0,183.63) (3.0,222.92) (4.0,241.65) (5.0,253.21) };
    \addlegendentry{pasta-poppy\(_\texttt{binary}\)};
    \addplot coordinates { (0.0,96.58) (1.0,115.05) (2.0,140.03) (3.0,190.56) (4.0,227.74) (5.0,251.77) };
    \addlegendentry{pasta-poppy\(_\texttt{SIMD}\)};
    \addplot coordinates { (0.0,104.5) (1.0,135.78) (2.0,176.77) (3.0,226.38) (4.0,247.51) (5.0,259.91) };
    \addlegendentry{pasta-poppy\(_\texttt{linear}\)};
    \legend{}
  \end{axis}
\end{tikzpicture}

\ref{leg:select_query_times_pasta_only}
\caption{Average select query time on all tested inputs of our implementations.}
\label{fig:select_query_times_pasta_only}
\end{center}
\end{figure}
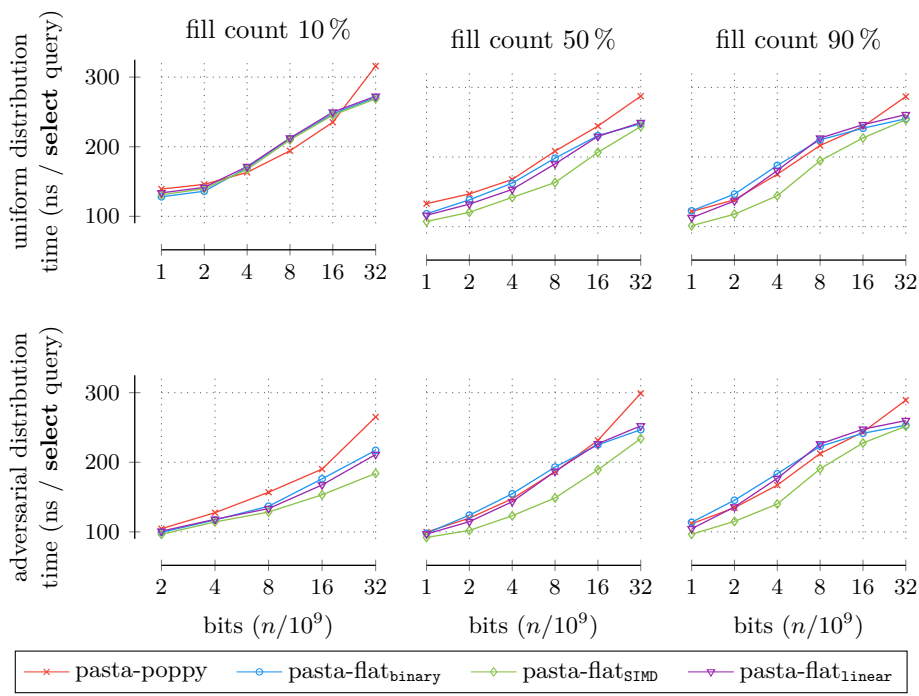

For better readability, we only show our fastest select algorithm in the main part of this paper.
Here, in \cref{fig:select_query_times_pasta_only}, we compare our implementations with each other.

\subsection{Construction Times}
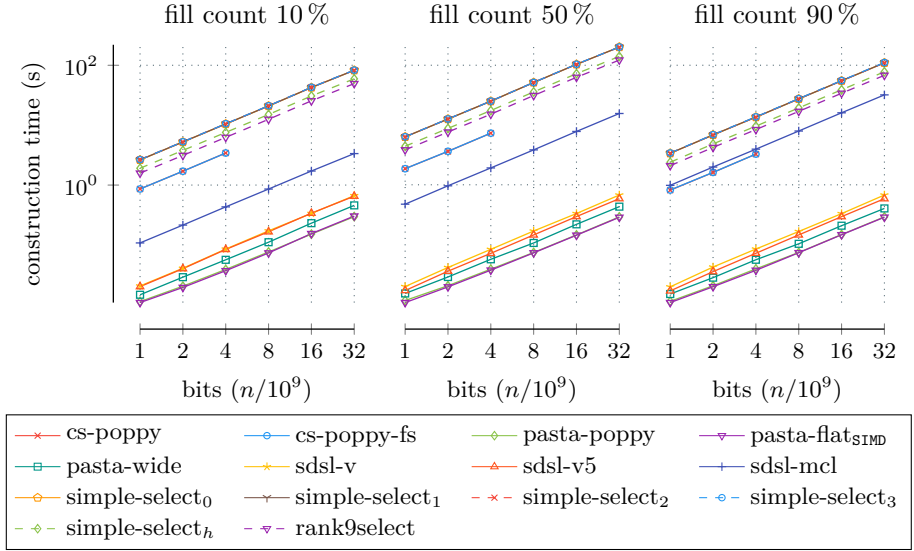
\begin{figure}[h!]
\begin{center}
\begin{tikzpicture}
  \begin{semilogyaxis}[
    plotConstructionTimeAll,
    legend columns=4,
    legend to name={leg:construction_times},
    title={fill count 10\,\%},
    xlabel={bits (\(n/10^9\))},
    ylabel={construction time (s)},
    ylabel near ticks,
    xlabel near ticks,
    xtick={0,1,2,3,4,5},
    xticklabels={1,2,4,8,16,32},
    ]

    \addplot coordinates { (0.0,0.8605) (1.0,1.70367) (2.0,3.43183) };
    \addlegendentry{cs-poppy};
    \addplot coordinates { (0.0,0.860667) (1.0,1.70433) (2.0,3.4275) };
    \addlegendentry{cs-poppy-fs};
    \addplot coordinates { (0.0,0.0113333) (1.0,0.0205) (2.0,0.039) (3.0,0.0763333) (4.0,0.151833) (5.0,0.297167) };
    \addlegendentry{pasta-poppy};
    \addplot coordinates { (0.0,0.0109167) (1.0,0.0195455) (2.0,0.0371818) (3.0,0.0735) (4.0,0.154833) (5.0,0.303833) };
    \addlegendentry{pasta-flat\(_{\texttt{SIMD}}\)};
    \addplot coordinates { (0.0,0.01475) (1.0,0.029) (2.0,0.0565455) (3.0,0.111167) (4.0,0.231167) (5.0,0.458) };
    \addlegendentry{pasta-wide};
    \addplot coordinates { (0.0,0.0206667) (1.0,0.0408333) (2.0,0.0853333) (3.0,0.172) (4.0,0.336167) (5.0,0.665333) };
    \addlegendentry{sdsl-v};
    \addplot coordinates { (0.0,0.0201667) (1.0,0.0401667) (2.0,0.0841667) (3.0,0.166) (4.0,0.3395) (5.0,0.652) };
    \addlegendentry{sdsl-v5};
    \addplot coordinates { (0.0,0.108167) (1.0,0.214833) (2.0,0.4325) (3.0,0.858) (4.0,1.71633) (5.0,3.35433) };
    \addlegendentry{sdsl-mcl};
    \addplot coordinates { (0.0,2.62983) (1.0,5.2315) (2.0,10.4632) (3.0,20.8893) (4.0,42.4682) (5.0,82.3925) };
    \addlegendentry{simple-select\(_0\)};
    \addplot coordinates { (0.0,2.63667) (1.0,5.24667) (2.0,10.488) (3.0,20.9327) (4.0,42.556) (5.0,82.4727) };
    \addlegendentry{simple-select\(_1\)};
    \addplot coordinates { (0.0,2.66983) (1.0,5.267) (2.0,10.5277) (3.0,21.0142) (4.0,42.797) (5.0,82.7608) };
    \addlegendentry{simple-select\(_2\)};
    \addplot coordinates { (0.0,2.65433) (1.0,5.278) (2.0,10.5507) (3.0,21.0663) (4.0,42.8958) (5.0,82.9405) };
    \addlegendentry{simple-select\(_3\)};
    \addplot coordinates { (0.0,1.908) (1.0,3.79) (2.0,7.574) (3.0,15.0853) (4.0,30.7202) (5.0,59.4232) };
    \addlegendentry{simple-select\(_h\)};
    \addplot coordinates { (0.0,1.5805) (1.0,3.1275) (2.0,6.2565) (3.0,12.5247) (4.0,25.4597) (5.0,49.4518) };
    \addlegendentry{rank9select};
  \end{semilogyaxis}
\end{tikzpicture}
\hspace{-.5cm}
\begin{tikzpicture}
  \begin{semilogyaxis}[
    plotConstructionTimeAll,
    title={fill count 50\,\%},
    xlabel={bits (\(n/10^9\))},
    y tick style={draw=none},
    yticklabel={\empty},
    y axis line style={draw=none},
    xlabel near ticks,
    xtick={0,1,2,3,4,5},
    xticklabels={1,2,4,8,16,32},
    ]

    \addplot coordinates { (0.0,1.859) (1.0,3.61583) (2.0,7.30917) };
    \addlegendentry{algo=efficient-poppy-rank-select};
    \addplot coordinates { (0.0,1.858) (1.0,3.634) (2.0,7.30933) };
    \addlegendentry{algo=efficient-poppy-rank-select-improved};
    \addplot coordinates { (0.0,0.0115833) (1.0,0.0205) (2.0,0.03875) (3.0,0.0736667) (4.0,0.145) (5.0,0.29) };
    \addlegendentry{algo=pasta-popcount};
    \addplot coordinates { (0.0,0.0107) (1.0,0.0195833) (2.0,0.0373) (3.0,0.0726667) (4.0,0.143333) (5.0,0.285667) };
    \addlegendentry{algo=pasta-popcount-flat-one\_dont\_care-intrinsics};
    \addplot coordinates { (0.0,0.0152) (1.0,0.0286667) (2.0,0.0567) (3.0,0.106333) (4.0,0.218) (5.0,0.430667) };
    \addlegendentry{algo=pasta-popcount-wide-one\_dont\_care-linear\_search};
    \addplot coordinates { (0.0,0.0198333) (1.0,0.0413333) (2.0,0.0831667) (3.0,0.1685) (4.0,0.3305) (5.0,0.6745) };
    \addlegendentry{algo=sdsl-rank-v};
    \addplot coordinates { (0.0,0.017) (1.0,0.0358333) (2.0,0.0713333) (3.0,0.144833) (4.0,0.294) (5.0,0.585833) };
    \addlegendentry{algo=sdsl-rank-v5};
    \addplot coordinates { (0.0,0.473833) (1.0,0.962) (2.0,1.91433) (3.0,3.83667) (4.0,7.805) (5.0,15.5807) };
    \addlegendentry{algo=sdsl-select-mcl};
    \addplot coordinates { (0.0,6.3235) (1.0,12.5693) (2.0,24.994) (3.0,51.4102) (4.0,103.457) (5.0,201.3) };
    \addlegendentry{algo=sux-SimpleSelect-0};
    \addplot coordinates { (0.0,6.33417) (1.0,12.5987) (2.0,25.0438) (3.0,51.3912) (4.0,103.437) (5.0,201.203) };
    \addlegendentry{algo=sux-SimpleSelect-1};
    \addplot coordinates { (0.0,6.33133) (1.0,12.6035) (2.0,25.0648) (3.0,51.4627) (4.0,103.5885) (5.0,201.298) };
    \addlegendentry{algo=sux-SimpleSelect-2};
    \addplot coordinates { (0.0,6.3485) (1.0,12.6223) (2.0,25.1068) (3.0,51.5338) (4.0,103.9775) (5.0,201.673) };
    \addlegendentry{algo=sux-SimpleSelect-3};
    \addplot coordinates { (0.0,4.41433) (1.0,8.774) (2.0,17.4572) (3.0,35.7767) (4.0,72.2143) (5.0,140.4195) };
    \addlegendentry{algo=sux-SimpleSelectHalf};
    \addplot coordinates { (0.0,3.83767) (1.0,7.621) (2.0,15.1533) (3.0,31.1197) (4.0,62.7458) (5.0,122.036) };
    \addlegendentry{algo=sux-rank9select};
    \legend{};
  \end{semilogyaxis}
\end{tikzpicture}
\hspace{-.5cm}
\begin{tikzpicture}
  \begin{semilogyaxis}[
    plotConstructionTimeAll,
    title={fill count 90\,\%},
    xlabel={bits (\(n/10^9\))},
    y tick style={draw=none},
    yticklabel={\empty},
    y axis line style={draw=none},
    xlabel near ticks,
    xtick={0,1,2,3,4,5},
    xticklabels={1,2,4,8,16,32},
    ]

    \addplot coordinates { (0.0,0.813667) (1.0,1.611) (2.0,3.26533) };
    \addlegendentry{algo=efficient-poppy-rank-select};
    \addplot coordinates { (0.0,0.8195) (1.0,1.61317) (2.0,3.26517) };
    \addlegendentry{algo=efficient-poppy-rank-select-improved};
    \addplot coordinates { (0.0,0.01125) (1.0,0.0205) (2.0,0.03925) (3.0,0.0738333) (4.0,0.147333) (5.0,0.290833) };
    \addlegendentry{algo=pasta-popcount};
    \addplot coordinates { (0.0,0.0107273) (1.0,0.01975) (2.0,0.0371818) (3.0,0.0731667) (4.0,0.146167) (5.0,0.287167) };
    \addlegendentry{algo=pasta-popcount-flat-one\_dont\_care-intrinsics};
    \addplot coordinates { (0.0,0.0149091) (1.0,0.028) (2.0,0.0558182) (3.0,0.103167) (4.0,0.207167) (5.0,0.402833) };
    \addlegendentry{algo=pasta-popcount-wide-one\_dont\_care-linear\_search};
    \addplot coordinates { (0.0,0.0196667) (1.0,0.0421667) (2.0,0.0843333) (3.0,0.165667) (4.0,0.329833) (5.0,0.6755) };
    \addlegendentry{algo=sdsl-rank-v};
    \addplot coordinates { (0.0,0.0168333) (1.0,0.0353333) (2.0,0.0716667) (3.0,0.145167) (4.0,0.293667) (5.0,0.585667) };
    \addlegendentry{algo=sdsl-rank-v5};
    \addplot coordinates { (0.0,0.980167) (1.0,1.98433) (2.0,3.9495) (3.0,7.94133) (4.0,16.0178) (5.0,31.9347) };
    \addlegendentry{algo=sdsl-select-mcl};
    \addplot coordinates { (0.0,3.37067) (1.0,6.82433) (2.0,13.5457) (3.0,27.3228) (4.0,54.8843) (5.0,109.601) };
    \addlegendentry{algo=sux-SimpleSelect-0};
    \addplot coordinates { (0.0,3.373) (1.0,6.83383) (2.0,13.579) (3.0,27.3658) (4.0,54.901) (5.0,109.68) };
    \addlegendentry{algo=sux-SimpleSelect-1};
    \addplot coordinates { (0.0,3.38433) (1.0,6.85467) (2.0,13.6012) (3.0,27.4715) (4.0,55.0658) (5.0,110.025) };
    \addlegendentry{algo=sux-SimpleSelect-2};
    \addplot coordinates { (0.0,3.40367) (1.0,6.8905) (2.0,13.6853) (3.0,27.5988) (4.0,55.4285) (5.0,110.768) };
    \addlegendentry{algo=sux-SimpleSelect-3};
    \addplot coordinates { (0.0,2.38633) (1.0,4.82917) (2.0,9.5785) (3.0,19.3383) (4.0,38.839) (5.0,77.7155) };
    \addlegendentry{algo=sux-SimpleSelectHalf};
    \addplot coordinates { (0.0,2.089) (1.0,4.21967) (2.0,8.38717) (3.0,16.9157) (4.0,33.962) (5.0,67.8517) };
    \addlegendentry{algo=sux-rank9select};
    \legend{};
  \end{semilogyaxis}
\end{tikzpicture}
\ref{leg:construction_times}
\end{center}
\caption{Average construction time in seconds over all inputs.}
\label{fig:construction_time}
\end{figure}

Finally, we want to report the construction times for the different data structures, see \cref{fig:construction_time}.
While the query times of the data structures are definitely more important, as we usually ask multiple queries but construct the data structure only once, we discovered that there are huge differences.
Overall, pasta-poppy, pasta-flat\(_\texttt{SIMD}\), and pasta-wide are the fastest to construct, followed by the data structures contained in the SDSL.
All other data structures (all variants of simple-select and cs-poppy) are orders of magnitude slower to construct.
The difference in construction time is so big, that we can answer more than 2\,000\,000 select queries using pasta-flat\(_\texttt{SIMD}\) and simple-select\(_2\) requires the same amount of time when we include the construction time.

\subsection{Answering Both Queries}
Note that our implementations are the only ones that can answer \(select_0\) and \(select_1\) without requiring twice the memory.\footnote{Per design, every rank-based select data structure can do so.}
Since all other select data structures in this evaluation are position-based, this is also the only data structure that can do so without increasing the space required by the data structure.
In \cref{tab:slowdown_select01}, we can see the slowdown of \(select_0\) queries compared with \(select_1\) queries, when the data structure is optimized for \(select_1\) queries.
Surprisingly, the binary search approach has the least slowdown.
Here, we have a trade-off: we can either double the required memory to answer both types of select queries or we can use a data structure that answers one of the queries 1.5 times slower.

\begin{table}[t]
  \centering
  \caption{Slowdown of \(select_0\) queries compared with \(select_1\) queries, when the ranks of ones are stored for each block. Note that our implementations also support optimized \(select_0\) queries.}
  \label{tab:slowdown_select01}
  \begin{tabular}{lrr}
    \toprule
    Name &  slowdown (uniform) & slowdown (adversarial)\\
    \midrule
    pasta-poppy & 1.632 & 1.559 \\
    pasta-flat\(_{\texttt{binary}}\) & \textbf{1.526} & \textbf{1.402} \\
    pasta-flat\(_{\texttt{SIMD}}\)  & 1.817 & 1.813 \\
    pasta-flat\(_{\texttt{linear}}\) & 1.651 & 1.604 \\
    \bottomrule
  \end{tabular}
\end{table}

\end{document}